\documentstyle[12pt,draft]{article}

\newcommand {\be} {\begin{equation}}
\newcommand {\bea} {\begin{eqnarray} \nonumber }
\newcommand {\ee} {\end{equation}}
\newcommand {\eea} {\end{eqnarray}}

\newcommand {\NE} {\not=}
 
\begin{document}

\title{P-adic numbers and replica symmetry breaking}
\author{ }
\maketitle

\vglue 1.5 true cm
\centerline{{\bf Giorgio Parisi$^a$ and Nicolas Sourlas$^b$}}
\medskip
\centerline{$^a$Dipartimento di Fisica and Sezione INFN,
Universit\`a di Roma ``La Sapienza''}
\centerline{Piazzale Aldo Moro, Roma 00199, Italy}
\centerline{e-mail: $\ $ giorgio.parisi@roma1.infn.it} 
\centerline{$^b$Laboratoire de Physique Th\'eorique de l' Ecole Normale
Sup\'erieure  \footnote 
{Unit\'e Propre du Centre National de la Recherche Scientifique,
associ\'ee \`a l' Ecole Normale Sup\'erieure et \`a l'Universit\'e de
Paris-Sud. } }
\centerline{ 24 rue Lhomond, 75231 Paris CEDEX 05, France. }
\centerline { e-mail: $\ $ sourlas@physique.ens.fr }

\begin{abstract}
The p-adic formulation of replica symmetry breaking is presented.
In this approach ultrametricity
is a natural consequence of the basic properties of the p-adic
numbers. Many properties can be
simply derived in this approach and p-adic Fourier transform seems to be an
promising tool.
 \end{abstract}
\newpage

\section{Introduction}

In the replica approach to disordered systems one usually introduces a 
matrix $Q_{a,b}$ which is the stationary point of a free energy $F[Q]$; the
matrix is a zero by zero 
matrix, with zero elements on the diagonal \cite{EA}.  Such a matrix is
constructed as the $n \to 0$ 
limit of a normal matrix with $n$ components.

In the mean field approach one looks for stable (or marginally stable) saddle
points of the free 
energy.  When the replica symmetry is spontaneously broken, as it happens in
spin glasses, one 
assumes that the saddle point is given by a matrix $Q$ constructed in a
hierarchical way, which 
corresponds to breaking the replica symmetry group (the permutation group of
$n$ elements) in a 
peculiar way \cite{mpv,parisibook2}.  The aim of this note is to expose some
hidden algebraic 
properties of this matrix and to show that the whole construction may be
simply done using p-adic 
numbers.

In this approach the ultrametric properties of the matrix $Q$ \cite{MPSTV,RTV}
arise naturally from 
the ultrametric properties of the p-adic numbers.  Although we do
not obtain new results 
in this way, we hope that this reformulation may be a useful starting point
for simplify some of the 
long computations involved in the evaluation of the corrections to the saddle
point approximation.

In section 2 will be present the basic properties of the p-adic construction
and show its 
equivalence to the usual hierarchical construction.  In the next section the
limit $n \to 0$ is 
performed in a simple way.  In section 4 we present an alternative and more
interesting procedure 
for doing the limit $n \to 0$, where we connect this approach to standard
p-adic analysis.  In the 
next section we show the advantages of using the p-adic Fourier transform. 
Finally there are two 
appendices, the first dedicated to the foundations of p-adic analysis, the
second to the basic 
properties of the Fourier transform\cite{ADIC, FOU}. 
 Both appendices can be skipped
by readers experts on 
p-adic analysis.

\section{The p-adic construction of the matrix $Q$}

We start the construction of the matrix $Q$ by considering a number $p$ (which
for simplicity we 
suppose to be a prime number) and by assuming that $n= p^L$ for some value of
$L$.  We are going to 
construct the matrix $Q$ for integer $L$ and $p$ in a specific way which we
will discuss later.  The 
limit $n \to 0$ will be done at the end.  The matrix $Q$ enters in the
evaluation of the free energy 
in spin glasses and related models in the saddle point approximation.  Here we
do not address the 
point of the evaluation of the free energy and we only consider the
construction of the matrix $Q$.

Eventually $n= p^L$ must go to zero and we can follow two
options in order to realise this goal:

a) We take a value of $p$ greater than one and we send $L$ to
minus infinity. Eventually we may
do an analytic continuation in $p$ to non integer values of $p$.

b) We first do an analytic continuation in $p$ to non integer values
of $p$.
 We take a value of $p$ less than one and we send $L$ to plus
infinity.

In both cases one obtain the limit $n\to 0$. The two constructions
are roughly equivalent. It
seems that the second one is more simple to work with, however
for pedagogical reasons we will start
by presenting the first one in section $3$, while the second one will
presented in section $4$.

The first steps are common to both strategies.  The construction of the
matrix $Q$ for integer 
$p$ and $L$ can be done as follows.  We assume that the matrix $Q_{a,b}$ is of
the form
\be
Q_{a,b} = Q(a-b).       \label{DIF}
\ee
where $Q(k)=Q(-k)$ (symmetric matrix) and $Q(k+n)=Q(k)$.  This choice
restricts very much the form 
of the matrix and shows an explicit symmetry of this parametrization (i.e.
translational invariance 
in internal space).  The condition $Q(0)=0$ implies that the elements on the
diagonal are equal to 
zero.

The second step consists in assuming that the function $Q(k)$ is a
function of
the p-adic norm $|k|_p$. The appendices provide a brief introduction to 
peadic analysis.

In other words we suppose that
\be
Q(k) = q(|k|_p).
\ee
This corresponds to setting
\be
Q_{a,b}= q_i \equiv q(p^{-i}),\ \ {\rm if} \ \ |a-b|_p = p^{-i}.
\ee

Before performing the limit $n \to 0$ it is convenient to compare
our apporach with the  standard hierarchical construction.

In the usual case  \cite{mpv} one introduces a sequence of $K+2$ numbers
$m_i$, with $m_0 = 1$ and $m_{K+1}=
n$, such that $m_{i-1}$ divides $m_{i}$ for $i=1,K+1$. One sets for $
a \NE b$:
\be
Q_{a,b}=q_i \ \ {\rm if} \ \ I(a/m_i) \NE I(b/m_i) \ \ {\rm and} \ \
I(a/m_{i+1}) = I(b/m_{i+1}),\label{HIE}
\ee
where the function $I(z)$ is the integer part of $z$, i.e. the largest
integer less or equal to $z$.

Let us consider the special case where the $m_i$ are given by
\be
m_i= p^i\label{POT}.
\ee
We want to show that the matrix $Q$ obtained in this way
coincides, after a permutation with the
matrix $Q$ constructed before with $K+1=L$.
The proof is rather simple. We associate to the index $a$ the $L$
digits of $a-1$ in base $p$:
\be
a= 1+ \sum_{i=0,K}a_i p^i
\ee
These digits form a $L$ dimensional vector with components in the
range $0\ -\ (p-1)$. The
hierarchical construction corresponds to set $Q_{a,b}=q_i$ if
$a_j=b_j$ for all $j \ge i$ and $a_{i-1}
\NE b_{i-1}$.

We now associate to an index $a$ its transpose ($a_T$), which is obtained
by writing its digits in the
inverse order:
\be
a_T= 1+ \sum_{i=0,K}a_i p^{(K-i)}.
\ee
The previous condition becomes that the $ K-i$ less significative digits of
$a_T$ and $b_T$ do 
coincide, and the $(K-i+1)^{\rm th}$ digit differs.  This last condition may
be restated by saying 
that $a-b$ is a multiple of $p^{(K-i)}$ but not of $p^{(K-i+1)}$, i.e.
$|a-b|_p=p^{-(K-i)}$.  Apart 
from a reshuffling of the indices, i.e. a permutation, the usual construction
is equivalent the 
p-adic construction presented before.

Generally speaking it is possible to prove that independently from the
condition in eq.  
(\ref{POT}), after a similar reshuffling of the indices, the hierarchical
matrix $Q$ (defined in eq.  
(\ref{HIE})) can always be written under the form of eq.  (\ref{DIF}).  It is
likely that many of 
the unexpected properties of the hierarchical construction arise from the
possibility of choosing an 
ordering of the indices in such a way that the hierarchical matrix is
invariant of under the 
transformation $a \to a+1$.  This invariance implies that the elements of one
line are the 
permutation of the elements of another line, however the converse is not true.

\section{The $n \to 0$ limit}

We can now perform the $n \to 0$ limit. We will firstly follow the
strategy a).

This limit can be
 reached by sending $L$ to $-\infty$. The continuation of the usual
formulae from positive to
negative $L$ can be done if we  introduce the quantities  $q_i$ for
non positive $i$.

 For
example let us consider the sum of the elements of a line of the
matrix. In order to perform the $n \to 0$ limit we slightly change the
notation of the
previous section and  we set
\be
Q_{a,b}= q_i ,\ \ {\rm if} \ \ |a-b|_p = p^{L-i}.
\ee
We easily get that the sum is given
\be
\sum_b Q_{a,b}=\sum_{i=1,L} (p-1)p^{i-1} q_i.
\ee
Indeed the number of integers $k$ such that $|k|_p\le p^{-j}$ (i.e. the volume
of the p-adic
sphere) is given by $p^{L-j}$ and therefore the number of integers $k$ such
that $|k|_p=p^{-j}$
(i.e. the volume of the p-adic shell) is given by $(p-1)p^{L-j-1}$.

 We are free to write the last equation it as
\be
 \sum_{i=-\infty,L} (p-1)p^{i-1} q_i -\sum_{i=-\infty,0} (p-1)p^{i-
1}q_i,
\ee
by introducing the extra parameters $q_i$ for $i<1$, which are
irrelevant for positive $L$.

 The
analytic continuation to negative $L$ can be now trivially done. In
the limit $L \to -\infty$, the
first term disappears and we get
\be
\sum_b Q_{a,b} \to \sum_{i=-\infty,0} (p^{i-1}-p^i) q_i.
\ee

A similar procedure can be followed in order to compute other
functions of the matrix $Q$.
By comparing the previous equation with the usual ones, we see
that we obtain the
hierarchical formulation where a function $q(x)$ is introduced, with
the extra constraint that
$q(x)$ is piecewise constant with discontinuities at $x=p^{-i}$. The
usual formulation, where $q(x)$
is a continuous function can be obtained by analytic continuation in
$p$ up to the point $p=1^+$.

By performing the explicit computations similar results are
obtained for the other quantities  and
it is possible to show that the usual approach is recovered.

\section{The upsidedown world}

In the other possible approach to the $n \to 0$ limit (b), we firstly do an
analytic continuation in 
$p$ to values less than one and only later we send $L \to + \infty$ in such a
way that $p^L \to 0$.  
At a later stage we are free to send $p \to 1^-$ in order to reach the
continuous limit.  In this 
way we get formulae quite similar to the previous one, with the advantage that
only the $q_i$ with 
positive $i$ are needed.

In this approach one obtains that the function $q(x)$ is given by
 \be
 q(p^i)= q_i,\label{DQ}
 \ee
where the index $i$ ranges from 0 to $+\infty$ in such a way that
when $p \to 1^-$, $x\equiv p^i$ spans the
interval 0-1. The formulae one obtains in this approach  for $p<1$
coincide with the formulae
obtained with the formalism of the previous section (with the
substitution of $p$ with $p^{-1}$).

 The advantage of this procedure is that we obtain formulae that
are very similar to those used
in the p-adic integral and that are well known to mathematician.
The strategy to prove these
formulae is quite similar and therefore one can use some of
the well known results in this
field.

In the region where $p<1$ it may be convenient to introduce the notation:
\be
|k|={|k|_p \over p^L }.
\ee
In this way $|k|$ belongs to the interval $0 \ - \ 1$,
with the exception $|0|=\infty$. In the limit where $p \to 1^-$, $|k|$
spans the interval 0-1. Equation \ref{DQ} can thus written as
\be
Q(a-b)=q(|a-b|).
\ee

Let us  apply this strategy to the computation of the sum of the
elements of a line of the
matrix. We find that
\be \lim_{L \to \infty}
\sum_{a=1,p^L} Q(a-b) = \sum_{i=1,\infty} (p-
1)p^{i-1} q_i .
\ee

For $p<1$ the previous equation can be written as
\be
 \sum_{a=1,n} Q(a-b) = \sum_{i=1,\infty} (1-p)p^{i-1} q_i ,
\ee
 while for $p>1$ the r.h.s. becomes proportional to the  p-adic integral which is
denoted as
\be
\int_p da Q(a).
\ee

With some abuse of notation we denote for $p<1$
\be
\lim_{n \to 0} \sum_{a=1,n} Q(a-b)= \int_p^{'} da Q(a),
\ee
where the sign ${'}$ over the integral $\int_p^{'}$ denotes that the value
zero is excluded from the 
integration range.  We 
must note that the measure of the integral is normalised to $-1$.  In a
similar way we can use the 
notation
 \be
\lim_{L \to \infty}  \sum_{b,c=1,p^L} F(|a-b|,|b-c|,|c-a|) =
 \int_p db \ da F(|a-b|,|b-c|,|c-a|).
\ee
For $p>1 $  we obtain the usual p-adic integral (apart from a
normalisation factor). The
results  for $p<1$ can be obtained using the same steps as in
appendix II.

We can do the computation in the interesting case where the sum is
restricted to all different
indices.
We have to compute
\bea
\lim_{L \to \infty}  \sum_{b,c=1,p^L;a\ne b,a\ne c,b\ne c} F(|a-b|,|b-c|,|c-a|)
\equiv \\
\lim_{L \to \infty}  \sum_{b,c=1,p^L} \ ^{'} F(|a-b|,|b-c|,|c-
a|)=\\
 \int_{p \ b,c=1,p^L;a\ne b,a\ne c,b\ne c}
db \ dc F(|a-b|,|b-c|,|c-a|)= \\
\int_p ^{'}da \ db F(|a-
b|,|b-c|,|c-a|), \nonumber
\eea
where we denote by $\sum \ ^{'}$ the sum restricted to the case of all
different indices.

In the same way we could define
\be
(-1)^M \lim _{n \to 0} {1 \over n} \sum_{a_1,a_2,...a_M} F[a] \equiv  \int
^{'} da_1 \ da_2 \ ... \  da_M,
F[a] \ee
where  $p$ is less than $1$ and the sum is done on all different indices.
The factor $(-1)^M $ has
the effect of giving a positive result for the integral.  Generally
speaking in this way the evaluation
of sums can be reduced to the computation of quantities that are
very similar to the corresponding p-adic
integral.

 For example let us use this strategy to compute

\be
\int ^{'}_p da \ db F(|a-b|,|b-c|,|c-a|).
\ee
 The application of the previous formulae tells us
that the integral is given by
 \bea
\sum_{b,c,b \NE a, c \NE a, b \NE c} F(|a-b|,|b-c|,|c-a|)=
\sum_{i,k,l}\mu(i,k,l)F(p^k,p^i,p^l)=\\
\sum_{i,k;i<k}
(p^{i+1}-p^i)(p^{k+1}-
p^k)[F(p^k,p^k,p^i)+F(p^k,p^i,p^k)+F(p^i,p^k,p^k)]+\nonumber\\
\sum_{i} (p^{i+1}-p^i)(p^{i+1}- 2 p^i)F(p^i,p^i,p^i).
\eea

The proof can be obtained using the same strategy as in the
appendix I for computing the measure of
three intersecting p-adic shells.

 Finally in the continuum limit where $p$ goes to $1^-$ one get the
formula \bea
\sum_{b,c,b \NE a, c \NE a, b \NE c} F(|a-b|,|b-c|,|c-a|)= \\
 \int dx dy \theta(x-y) [F(x,x,y) + F(x,y,x) +F(y,x,x) ]+ \int dx \ x \
F(x,x,x).
\eea
The same formula could be simply written as
\bea
\sum_{b,c,b \NE a, c \NE a, b \NE c} F(|a-b|,|b-c|,|c-a|)=\int dx dy dx
\mu(x,y,x) F(x,y,z) \\
\mu(x,y,z)= 
\theta (x-y) \delta(x-z) +\theta (x-z) \delta(x-y)
+\theta (y-x) \delta(y-z) +
 x\delta(x-y) \delta (x-z) 
\eea

It is important to note  that the ultrametricity inequality works at
reverse in the region $p < 1$
and consequently also in the limit  $p \to 1^-$. This is in agreement
with the fact that $1-x$, not $x$,
has the physical meaning of distance.

 After some work one can find simple rules for generic
sums of the type
\be {
-1 \over n}\sum_{a,b,c,d}\ ^{'} F(a,b,c,d) =  \int _p^{'} F(a,b,c,d)
\ee
where $F$ depends only on the p-adic distance and all indices are
different \cite{EP}.
In the case where the function is symmetric one finds that in the
limit $p \to 1^-$
\bea
&\int ^{'}_p F(a,b,c,d) = \\
&\int_{x<y<z} dx \ dy \ dz  F|_{|a-b|=|a-c|=|a-d|=x,|b-c|=|b-d|=y,|c-
d|=z} + 11\  permutations +
\nonumber \\
&\int_{x<y;x<z} dx \ dy \ dz  F|_{|a-b|=z,|b-c|=|b-d|=|a-c|=|a-d|=x,|c-
d|=y} + 2\  permutations
\nonumber\\
&+\int_{x<y} dx \  dy \ x \ F|_{|a-b|=y,|b-c|=|b-d|=|a-c|=|a-d|=x=|c-
d|=x} +5\  permutations
\nonumber \\
&+\int_{x<y} dx \  dy \ y \ F|_{|a-b|=|b-c|=|a-c|=y,|b-d|=|a-d|=x=|c-
d|=x} +3\  permutations
\nonumber \\
&+ 2! \int dx \ x^2 F(_{|a-b|=|a-c|=|a-d|=|b-c|=|b-d|=|c-d|=x}),
\eea
where the formula for the intersection of 4-p-adic sphere of the
same radius has been crucial to
obtain the last term.

 If we apply the same strategy to more complicated sums we can
find the formula of ref.
\cite{MY}, where the result is written as sum over all possible trees,
with a specific integral
associated to a given tree.

\section{Using the p-adic Fourier Transform}
An interesting application of this approach can be done to  the
formula for the product of two
matrices $A$ and $B$:
\be
C_{i,k}=\sum_j A_{i,j}B_{j,k}
\ee
If the matrices have the form  discussed here one finds that the
previous formula can be written as
a convolution
 \be
C(i-k)=\sum_j A(i-j) B(j-k)=\sum_j a(|i-j|) b(|j-k|).
\ee
Finally one finds using the previous formulae that
\be
c(|i|)= \int_p dk \ a(|i-k|)b(|k|) ,
\ee
which using the rules of p-adic integral after performing the limit
$p \to 1^-$ can be written as:
\be
c(x) = \int_x^1 dy (a(y) b(x)+ a(x)b(y))+ \int_x^1 dy a(y) b(y) + x \
a(x)b(x).\label {PROD}
\ee
Convolutions can be strongly simplified in Fourier space.
In principle we can just do ordinary Fourier transform, where the
momentum $q$ is in the interval
$(-\pi\ , \ \pi)$, however it is convenient to take into account the
p-adic nature of the functions
we consider.

We can start from the analysis leading to formula (\ref{FOU}) of appendix
(II).  Generalising the 
computations to the case where $p<1$ and performing the continuum limit, one
finds that the p-adic 
Fourier transform\footnote{We use the same notation as the appendix and we use
the square 
parenthesis, 
$[\cdot]$, to denote the Fourier transform} of a function $A(k)=a(|k|)$ is a
function $A[M]=a[y]$, 
defined for $y$ in the interval 0-1, where $y=|M|^{-1}$.  One must be careful
in removing the 
factor ${1 \over p^L}$ in the normalization of the Fourier transform, which
would be harmful here.

Eq.  (\ref{FOU}) may be transformed to
\be
A[M]|_{|M|_p=p^j}=
 \sum_{k=0,j}\left( p^{-k+L} (1 -1/p) a_{L-k}\right)- p^{-j+L} a_{L-
j}+ A(0).
\ee
Let us define (for $p<1$) the function
\be
a[y]=A[M]|_{|M|_p=p^L y}
\ee
The function $a[y]$ is defined only for $y$ of the form $p^{-k}$ for integer $k$.
With this definition one finally obtains that
\be
a[y]=\sum_{k=0,j} p^{k} (1 -1/p) a(p^{-k}) - p^{j} a(p^{- j})+ A(0).
\ee
The final formulae in the continuum limits are
 \bea
 a[y] = a(\infty)-  \int_y ^1 dx a(x) - y a(y), \\
 a[\infty]= a(\infty) - \int_0 ^1 dx a(x),\label{F1}
\eea
where we use the apparently strange notation  $a(\infty)= A(0)$ and
$a[\infty]= A[0]$.

The inverse Fourier transform formulae simply read
\bea
a(x)=  a[{\infty}]-a[0] + \int_0^x {dy \over y} {da[y] \over dy},\\
a(\infty)= a[1].\label{F2}
\eea

These relations becomes simpler in differential form. For example
one gets:
\be
x\ a'(x)= a'[x] .\label{FOUDIF}
\ee
This differential relation is equivalent to the integral relation in
eq. (\ref{F1}, \ref{F2}) for obtain the Fourier transform if they are
complemented by value of the Fourier transform in a given point. A
possible choice is
\bea
a[1]=a(\infty) -a(1).\label{F3}
\eea

With some work one can
 verify that the multiplication of two  matrices becomes the simple
multiplication of their Fourier transform \cite{MP}:
\be
 c[y] =  a[y] \ b[y].
\ee
Indeed diffentiating eq. (\ref {PROD}) one finds
\be
c'(y) = a'(x)(\int_x^1 dy b(y) + x b'(x))+ b'(x)(\int_x^1 dy a(y) + x a'(x))
\ee

The function $ a[y]$ was already introduced in ref. \cite{MP} in
order to solve the
inversion problem, although its p-adic nature was not recognised. Also the
usual procedure of
simplifying   the saddle point equations by differentiating them correspond to
consider the
p-adic Fourier transform.

 We could also consider
the problem of computing the inverse of a  matrix $Q$, i.e. of finding
matrix $R$, such that
\be
\sum_c Q_{a,c} R_{c,b} = \delta _{a,b}.
\ee
It is not a surprise that we find the Fourier transform of the matrix
$R$ is simply the inverse of
the Fourier transform of the matrix $Q$:
\be
 q[y] =1/ r[y].
\ee

An extremely important problem consists in the computation of the
inverse of the Hessian coming
from the fluctuation around the saddle point. Here one has to solve
the equation
\be
\sum_{e,f} M_{a,b;e,f} G_{e,f;c,d} = \delta_{a,b;c,d}.
\ee
This inversion is not
a simple job and rather complex computations have been done
\cite{DK1,DK2}. However
the final formulae are remarkable simple. Although we are not able at
the present moment to derive these formulae in the framework of the
p-adic formalism it may be useful to  show that they have a very simple
interpretation in terms of p-adic Fourier transform \cite{FOU}.

We will consider here only the so called replicon sector for which
the results are simpler than in the other sectors.
 We restrict our analysis to the region where $|a-b| = z>
|a-c|=x_1, |b-d|=x_2>z$. In  this region ultrametricity
implies that $z=|a-d|=|b-c|=|c-d|$. Both $M$ and $G$ are functions of
 $x_1,x_2,z$ only and we write
them as $M^z(x_1,x_2)$ and $G^z(x_1,x_2)$. In the same way we
denote by $G_R^z(x_1,x_2)$ the
replicon contribution to $G$, where the precise definition of the
replicon can be found in the
original papers \cite{DK1,DK2}.

Following \cite{DK1} we can thus introduce the  Fourier
transform with respect to $x_1$ and
$x_2$, which is given by
 \bea
m^z(x_1,y_2] =  m^z(x_1,\infty)- \int_{y_2}^1 dx_2 \ m^z(x_1,x_2)
-y_2 m^z(x_1,y_2),\\
m^z[y_1,y_2] =  m^z(\infty,x_2]- \int_{y_1}^1 dx_1 \ m^z(x_1,y_2]
-y_1 m^z(x_1,y_2].
 \eea

One finally finds that the final formula for the replicon
propagator \cite{DK2} may be obtained with slightly modified
inverse Fourier Transform:
 \bea
g_R^z[x_1,x_2]={1 \over m^z[x_1,x_2]},\\
x_1 x_2 {\partial^2 \over \partial x_1 \partial x_2
}g_R^z(x_1,x_2)={\partial^2 \over \partial x_1 \partial x_2
}{1 \over m^z[x_1,x_2]},\\
g_R^z(z,x_2)=g_R^z(x_1,z)=0. \nonumber
\eea

Equivalently we could write the last equation as
\be
g_R^z(x_1,x_2)=\int_{x_1}^z dx_1 \int_{x_2}^z dx_2
{1 \over {x_1 x_2}}{\partial^2 \over \partial x_1 \partial x_2
}{1 \over m^z[x_1,x_2]}.
\ee
The differential relations in the inverse Fourier transform
are preserved, only the second condition which fixes the value of the inverse
Fourier transform in one point is modified. With these modifications the
replicon sector of the inverse is just the numerical inverse of the
matrix $M$ in   Fourier space.

The precise reason  for the appearance of these simple
formulae with a strong p-adic flavour is not completely clear at  the
present moment. They show the usefulness of the p-adic formalism.
It would be also extremely interesting to study if the same formalism could be
applied to the off equilibrium dynamics of the kind studied in ref \cite{CK}.

After completion of this work we received a paaper by 
V. A. Avetisov, A. H. Bikulov, S. V. Kozyrev \cite{abk} where some similar results 
are derived. 

\section*{Appendix I: p-adic numbers}

Let us consider a prime number $p$. Any integer $k$ can be written in an
unique way  as
\be
 k= p^i \sum_{l=0}^{\infty}  a_l p^l
\ee
with $i \ge 0 $, $0 \le a_l \le p-1 $ and $ a_0 \ne 0 $.
The p-adic norm of such an integer
$k$ (i.e. $|k|_p$) is defined as
\be
|k|_p = p^{-i},
\ee
 The p-adic norm of 0 is defined to be equal to
zero. The value of the p-adic norm tells us the number of
consecutive zeros at the end of a number,
when it is written in base $p$. For a rational number $ r= a/b$, the p-adic
norm is defined as $ |r|_p = |a|_p / |b|_p $.

The properties of the p-adic norm are well studied by
mathematicians, one of the most famous
property being ultrametricity, which states that
\be
|a-b|_p \le \max(|a-c|_p,|c-b|_p)
\ee
for any choice of $c$. This property, which generalises the
statement {\sl the sum of two even
number is even}, can be proved as follows.

Using the translational invariance of the metric, we first write the
ultrametric inequality in an 
equivalent way as
\be
|a+b|_p \le \max(|a|_p,|b|_p).
\ee
If $a$ is a multiple of $p^i$ (and not of $p^{i+1}$), and $b$ is a
multiple of $p^k$, with $k \ge i$,
it is evident that $a+b$ is a multiple of $p^i$. Therefore
\be
|a+b|_p \le p^{-i} = |a|_p =\max(|a|_p,|b|_p).
\ee
We stress that we have used in a crucial way the fact that $p>1$ for
a {\sl true} prime. (In this paper we make an analytic continuation to
$p<1$. In that case, the inequality sign would be reversed.)
A direct consequence of this inequality is that any triangle is either
equilateral or isosceles with the two largest sides equal. It follows
that any point $a$ inside the p-adic disk centered at $o$ and of radius $r$,
i.e. such that $ |a-o|_p \le r $, is also a center of the disk;
i.e. if  $ |b-o|_p \le r $, then also $ |a-b|_p \le r $.

The whole p-adic field may be constructed starting from the p-adic rationals by
considering the closure of the rationals with respect to the p-adic
norm in the
same way that the real numbers (of the interval $0-1$) are
constructed as the closure of the
rationals (of the interval $0-1$) with respect to the usual
Euclidean norm.

Closing the rational field with respect to the previously defined norm
one obtains the p-adic field.
Continuity of a p-adic function can be defined as usual. For example
a  function $f$ is
continuous at the point $k$ if \be
\lim_{n \to \infty} f(k_n) = f(k),
\ee
for any sequence of $k_n$ which converges to $k$ in p-adic sense (i.e.
$|k_n-k|_p \to 0$).  The 
extension of a function from integers to p-adic numbers is called p-adic
interpolation.  Here we do 
not need to discuss this point any more.

>From our point of view a more interesting construction is the
integral over the p-adic integers
which can be defined in an elementary way as
 \be
\lim_{L \to \infty}
{1 \over p^L} \sum_{a=1}^{p^L} F(a) = \int_p da F(a).
\ee

 There are many well known properties of the p-adic integral. Here
we report some of them, leaving the proof to the reader (the lazy
reader can found them in any book
on p-adic calculus).

a) The measure of the p-adic sphere of radius $p^{-i}$ centred
around an arbitrary point $a$  (i.e.
the measure of all points such that $|a-b|_p\le p^{-i}$) is given by
$p^{-i}$.
As far as the p-adic distance among integers
cannot be larger than 1, the unit sphere coincides with the whole
space and has measure 1.

b) The measure of the p-adic shell of radius $p^{-i}$ centred around
an arbitrary point $a$
(i.e. the
measure of all points such that $|a-b|_p= p^{-i}$) is given by $p^{-i}
- p^{-i-1} = (1-p^{-1})
p^{-i}$.

c) The measure of the intersection among two p-adic shells has
rather interesting properties. Let us
consider the intersection of a shell of radius  $p^{-i}$ centred
around the point $a$  with a shell
of radius $p^{-k}$ centred around the point $b$. The measure
depends on the distance among the points
$a$ and $b$,  which we assume to be equal to $p^{-j}$.
Ultrametricity tell us that the measure is
zero unless two among the distances coincide and the two equal
distances are the largest. After some
reflection one finds that only three cases have to be
considered.
\begin{itemize}
\item
We first consider the case $i=k < j$. Here the ultrametricity
inequality implies that the two
shells coincide and therefore the measure of the intersection is
simply given by  $(1-p^{-1})
p^{-i}$.
 \item
We now consider the case $i=j< k$. Here the ultrametricity
inequality implies that the second
shells is fully contained in the first one and therefore the measure
of the intersection is simply
given by  $(1-p^{-1})p^{-k}$.
\item
We finally consider the less trivial case is $i=j=k$. If one notice that
the two spheres of radius
$p^{-i}$ centred in $a$ and $b$ coincide and that the two spheres
of radius
$p^{-i-1}$ centred in $a$ and $b$ have zero intersection one finds
that the measure of the
intersections of the two shells is given by $(1-2p^{-1})p^{-i}$,
 \end{itemize}

d) The generalisation of the previous arguments allows us to compute the
measure of the intersection 
of many p-adic shells by using ultrametricity in a systematic way.  The most
significative result is 
that the intersection of $M$ shells of radius $p^{-i}$, whose centres are all
at mutual distance 
$p^{-i}$ is given by $(1-Mp^{-1})p^{-i}$.  The measure becomes zero for $p=M$,
which implies that 
you cannot find $M+1$ numbers exactly at the same distance.  This last results
is a generalisation 
of the well known statement that you cannot find three integers ($a, b, c$)
such that the three 
differences among them ($a-b,b-c,c-a$) are all odd.

  Using the previous formulae
there are a few p-adic integrals that can be obtained  a simple way.

 For example let us
try to compute
\be
\int_p da \ db F(|a-b|,|b-c|,|c-a|),
\ee
where for simplicity we denote by $|a|$ the p-adic norm of $a$.

The integral is $c$ independent and the application of the previous
formulae tells us that the integral is given by
 \bea
&\sum_{i,j,i\ne j} p^{-i} p^{-j} (1-p^{-1})^2 F(p^{-k},p^{-i},p^{-j})+\\
&\sum_{i,j > i}  p^{-i} p^{-j} (1-p^{-1})^2 F(p^{-j},p^{-i},p^{-i})+\\
&\sum_{i}  p^{-2i}(1- 2 p^{-1})(1- p^{-1}) F(p^{-i},p^{-i},p^{-i}),
\eea
where $k=min(i,j)$

\section*{Appendix II: p-adic Fourier transform}

 Fourier transform on the p-adic integers coincides with the usual Fourier transform.
It can also be defined by analysing the characters of the
addictive group. It is more simple to consider first the case where
$L$ is
finite and only a finite number of points is present.

        We start by considering the case in which the function $A(x)$
is defined only for $x=1, \cdots, p^L$ (with $A(0)=A(p^L)$). The
Fourier transform is defined as
\be
A[M] ={1 \over p^L} \sum_{m=1}^{p^L} \exp (2 \pi i  m M) A(m),
\ee
where $M$ is a rational number of the form $j p^{-L}$ with $0 \le j
< p^L$. As usual the Fourier
space contains the same number of points of the original space. In
this paper we will use the square
parenthesis to denote Fourier transform.

Let us consider the problem of computing the Fourier transform of
a function which depends only on
the p-adic norm,i.e. $A(k)=a(|k|_{p})$. We are thus interested in computing

\be
A[M] = {1 \over p^L} \sum_{m=1}^{p^L}   \exp (2 \pi i \ m M) \ \ \
a(|m|_p)=
\sum_{k=0}^{L-1} a(p^{-k}) S_k[M] +{1 \over p^L} A(0),
\ee
where $S_k[M]$ is the Fourier transform of the p-adic shell of
radius $ p^{-k}$:
\be
S_k[M]= {1 \over p^L} \sum_{m=1}^{p^L} \delta_{|m|, p^{-k}} \exp
(2 \pi i \ m M).
\ee

In order
to compute the Fourier transform of the shell, it may be simpler to
firstly compute the Fourier
transform of the p-adic sphere of radius $ p^{-k}$. A simple
computation shows that
\bea V_k[M]= {1
\over p^L} \sum_{m=1}^{p^L} \theta( |m|-  p^{-k}) \exp (2 \pi i \ m M) =
 {1 \over p^{L}} \sum_{m=1}^{p^{L-k}}  \exp (2 \pi i \ m M p^k) = \\
= \ p^{-L} \exp ( 2 \pi i l  p^{k-L} )  \
{ { \exp ( 2 \pi i l ) -1  \over
{ \exp ( 2 \pi i l  p^{k-L} }) -1 }  }, 
\eea
where we used the definition $M =  l p^{-L} $.

It follows that $ V_k[M]= 0 $ unless $ l = n p^{L-k} $, $n=0,1,\cdots,p-1 $
and the last passage is 
no more valid, because both numerator and denominator are equal to zero in the
final result.
We notice that the possible values of $|M|$ are $p^j$ for non
negative $j$. Consequently we find that  $ V_k[M]= 0 $ unless $ k- j \le 0$
We also remark that, as consequence of translational invariance, the Fourier
transform of a function of the
p-adic norm, is still a function of the p-adic norm.
We can thus define the functions $a[ \ ]$ and $s_k[ \ ]$ as
\bea
a[p^{-j}] = A[M]\ |_{|M|=p^j}\\
s_k[p^{-j}] = S_k[M]\ |_{|M|=p^j}
\eea
The reader should notice that the functions $a[\ ]$ and $s_k[\ ]$ are
defined in such a way that it
argument is the range $0-1$. It follows that

\bea
v_k[p^{-j}] = p^{-k}  &{\rm for}& k -j \le 0\\
v_k[p^{-j}] = 0 &{\rm for}& k -j> 0.
\eea

As a consequence we find that the Fourier transform of a spherical
shell is  given by
\bea
s_k[p^{-j}] = p^{-k} (1 -1/p) &{\rm for}& k -j < 0,\\
s_k[p^{-j}] = -p^{-k}  &{\rm for}& k -j = 0, \\
s_k[p^{-j}] = 0 &{\rm for}& k -j > 0 \nonumber.
\eea

  The final
expression for the Fourier transform thus
becomes \be
 a[p^{-j}] = \sum_{k=0,j}p^{-k} (1 -1/p) a(p^{-k}) + p^{-j} a(p^{-
j})+ {1\over p^L} A(0).\label{FOU}
\ee

It is interesting to note that the last formula the dependance on $L$
is very simple, so that the limit $L
\to \infty$ can be trivially done. Moreover most of the properties
of the ordinary Fourier
transform, like the theorems concerning convolutions, are still valid.


\begin{thebibliography}{99}

\bibitem {EA} S.F.Edwards and P.W. Anderson, J. Phys. F5, 965
(1975).

\bibitem{mpv} For a review see M.~M\'ezard, G.~Parisi and
M.A.~Virasoro, {\sl Spin glass theory
and beyond} (World Scientific 1987).

\bibitem{parisibook2} G.~Parisi, {\sl Field Theory, Disorder and
Simulations},
(World Scientific 1992).

\bibitem {MPSTV}  M.~M\'ezard, G. Parisi, N. Sourlas, G. Toulouse
and M. Virasoro, J. Phys. 45,
843(1984).

 \bibitem {RTV}R. Rammal, G. Toulouse and M.A. Virasoro, Rev.
Mod. Phys. 58, 765 (1986).

\bibitem {ADIC} M. H. Taibleson {\sl  Fourier Analysis on local Fields},
Princeton University Press (Princeton, NJ), 1975.


\bibitem{FOU}   C. De Dominicis, D. M. Carlucci and T. Temesvari, J. Phys. I
France {\bf 7}, 105 (1997);
D. M. Carlucci and C. De Dominicis {\sl On the Replica Fourier Transform} cond-mat/9709200.

\bibitem {MY}  M.~M\'ezard  and  Yeidia (unpublished).

\bibitem {EP} E. Marinari and G. Parisi, Phys. Lett. 203, 52 (1988).
\bibitem {MP}  M.~M\'ezard and G. Parisi J. Phys. I France 2,
2231 (1991).

\bibitem {DK1} C. De Dominicis and I. Kondor, Europhys. Lett. 2, 617.
(1986)
\bibitem {DK2} C. De Dominicis and I. Kondor, J. Physique Lett. 46,
L1037 (1985).
\bibitem {CK} L.F. Cugliandolo and J. Kurchan. Phys. Rev. Lett. 71, 1
(1993).
\bibitem {abk} V. A. Avetisov, A. H. Bikulov, S. V. Kozyrev, cond-mat/9904360
\end{thebibliography}
 \end{document}